# Optical microscope based universal parameter for identifying layer number in two-dimensional materials

Mainak Mondal, Ajit Kumar Dash, Akshay Singh*

Department of Physics, Indian Institute of Science, Bengaluru, India- 560012

*Corresponding author: aksy@iisc.ac.in

**Abstract:**

Optical contrast is the most common preliminary method to identify layer number of two-dimensional (2D) materials, but is seldom used as a confirmatory technique. We explain the reason for variation of optical contrast between imaging systems. We introduce a universal method to quantify the layer number using the RGB (red-green-blue) and RAW optical images. For RGB images, the slope of 2D flake ($MoS_2$, $WSe_2$, graphene) intensity vs. substrate intensity is extracted from optical images with varying lamp power. The intensity slope identifies layer number and is system independent. For RAW images, intensity slopes and intensity ratios are completely system and intensity independent. Intensity slope (for RGB) and intensity ratio (for RAW) are thus universal parameters for identifying layer number. A Fresnel-reflectance-based optical model provides an excellent match with experiments. Further, we have created a MATLAB-based graphical user interface that can identify layer number rapidly. This technique is expected to accelerate the preparation of heterostructures, and fulfil a prolonged need for universal optical contrast method.

**Introduction:**

Two-dimensional (2D) materials consist of single or few atomic layers of materials with remarkable optoelectronic properties[1,2]. There is vast potential of 2D materials for sensing[3–5], quantum computing[6,7], and study of moiré physics[8,9]. Single or few-layer samples can be prepared through top-down (mechanical[10] and chemical exfoliation) or bottom-up (chemical vapor deposition[11]) methods. Mechanical exfoliation is the most common process for creating high-quality 2D material flakes. However, exfoliation produces randomly distributed flakes over the substrate, with varying layer number. The layer number is identified by various methods, including Raman spectroscopy, atomic force microscope (AFM), photoluminescence (PL), or optical contrast. Raman spectroscopy, AFM, or PL setups consist of sophisticated machinery and dedicated systems. The layer identification methods thus make the device preparation process slower and more costly, especially for multi-layer stacked samples[12]. On the other hand, the optical contrast method only needs a simple optical microscope imaging system, making this method highly efficient and low cost.

Numerous studies discuss the identification of layer number by just using optical microscope images. The changes between substrate intensity ($I_{Sub}$) and the 2D material flake intensity ($I_F$) for individual red, green, blue channels, or the average of these channels can identify different layered regions[13,14]. Specifically, the contrast difference ($C_D$), $\frac{(I_{Sub}-I_F)}{(I_{Sub}+I_F)}$ or $\frac{(I_{Sub}-I_F)}{I_{Sub}}$, measured for different channels, is used for layer identification[15–17]. Most of the reported $C_D$ parameters are based on the intensity ratio (α) between $I_F$ and $I_{Sub}$ ($C_D = \frac{1-\alpha}{1+\alpha}$ or $1-\alpha$). In these studies across different labs and imaging systems, separate imaging conditions are maintained. As a result, optical contrast is only used for quick identification purposes instead of final confirmation.

This study aims to identify an easily accessible and universal parameter to measure layer number. We image mechanically exfoliated $MoS_2$ flakes (on 285 nm $SiO_2$/Si substrate) using different imaging systems, in RGB and RAW formats. In the case of RGB images, we find that α can identify layer number within the same imaging system, but values of α (for a fixed layer number) vary across systems. Also, there is significant change in α for images taken with different lamp powers. We realized that the unavoidable post-processing effects introduced during digital RGB image formation cause such variations in α. We found that $I_F$ and $I_{Sub}$ vary linearly (for a large intensity range) for the images taken with increasing light intensities (for both formats) and the calculated intensity slopes (μ) vary with layer number. Remarkably, the μ values for RGB format remained consistent for different imaging systems. For the RAW format, μ (same as α) were found to be completely independent of microscope systems and the lamp power. A Fresnel-reflectance-based imaging model is used to calculate and confirm the measured α values based on RAW image formation. Hence, we suggest that intensity slope (for RGB) and intensity ratio (for RAW) can be used universally to identify layer number with high confidence. We extended the technique to graphene and $WSe_2$ as well with similar consistency. Finally, we have created a MATLAB-based GUI to check the slope values and confirm different layered regions quickly, allowing wide adoption of this method.

**Results:**

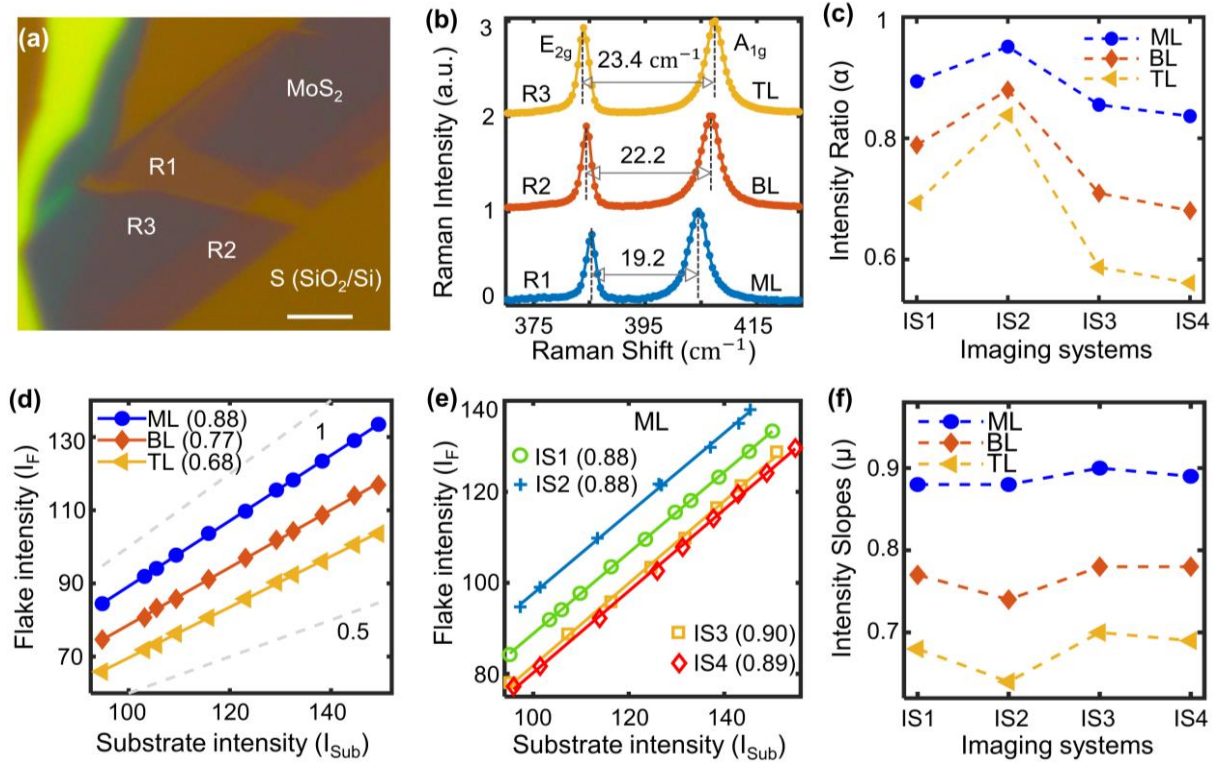

**Figure 1 (Intensity ratio and slope analysis for MoS$_2$ RGB images taken using different Imaging systems): (a)** Optical image of mechanically exfoliated MoS$_2$ on a 285 nm SiO$_2$/Si substrate. Regions with different colour contrasts are labelled as S (substrate), R1, R2, and R3. The scale bar is 5 μm. **(b)** Raman peak separation of E$_{2g}$ and A$_{1g}$ vibrational modes for R1, R2, and R3 identifies the regions as monolayer (ML), bilayer (BL), and trilayer (TL), respectively. **(c)** Variation of red channel intensity ratios (α) measured using four different imaging systems (IS1, IS2, IS3, and IS4). **(d)** Variation of the reflected flake (ML, BL, TL) intensity (I$_F$) with substrate intensity (I$_{Sub}$) for increasing lamp light power. Intensity slopes (μ) of each line are indicated in labels. Lines of slopes 1 and 0.5 are drawn to improve visualization. **(e)** I$_F$ vs. I$_{Sub}$ for ML, measured using the different imaging systems. The labels are written in the format: imaging system (slope). **(f)** μ for ML, BL, TL regions were measured using the different imaging systems and showed consistency across different imaging systems.

| Imaging System # | Microscope model | Light source | Camera |
|---|---|---|---|
| IS1 | Leica – DM2500M | Halogen | Leica CMOS - DFC400 |
| IS2 | Olympus – BX51 | Halogen | Olympus CCD – UC30 |
| IS3 | Custom built | Halogen | Amscope CMOS – MU1803 |
| IS4 | Olympus – BX53M | White LED | Amscope CMOS – MU1803 |

**Table 1:** Component details for the different microscope imaging systems used in this study.

We start with mechanically exfoliated $MoS_2$ on 285nm $SiO_2$/Si substrate, imaged by imaging system IS1 (see Table 1 for details on imaging systems), shown in Figure 1a. Regions with different optical contrast are expected to be regions with different layer number (labelled as R1, R2, and R3, substrate labelled as S). To identify the layer number, Raman spectroscopy is performed (shown in Figure 1b). The peak separation of the longitudinal ($E_{2g}$) and out-of-plane ($A_{1g}$) vibrational mode characterizes the regions R1, R2, and R3 as monolayer (ML), bilayer (BL), and trilayer (TL), respectively[18,19].

The intensity ratios (α) are calculated from the substrate ($I_{Sub}$) and flake intensity ($I_F$) for the red, green, and blue channels. The procedure of extracting these intensity values is described in Supplementary Figure S1. For $MoS_2$, the calculated reflectance difference between flake and substrate regions is highest in the 550-700 nm range, corresponding to the red channel[15] (Supplementary Section 6). As a result, images from the red channel show the most noticeable changes in α for different layer-numbered regions. Comparison of α for different channels is shown in Supplementary Figure S1. The red channel α for ML, BL, and TL $MoS_2$ flake (shown in Figure 1a) corresponding to images taken using four different microscope imaging systems, are plotted in Figure 1c. Change of the imaging system causes significant variations in α values. This will lead to incorrect identification of layer number. For example, α measured for ML using IS4 is the same as α measured for BL using IS2. Even for the same system, α has considerable dependence on imaging light intensity (discussed later, Figure 3b). Hence α and $C_D$ parameters cannot be used universally to identify layer number.

Next, images of the sample are taken at different lamp powers using the imaging system IS1. $I_F$ is found to vary linearly with $I_{Sub}$, with the three different layer-numbered regions having three different μ (Figure 1d). Note that these μ values are quite distinct and can differentiate between different layer-numbered regions. Next, another set of similar images is taken using IS2, IS3, and IS4. Remarkably, μ values are the same for ML, independent of imaging systems (Figure 1e, substrate intensity range 95 to 160). We also note that the intercepts are different for all the microscope systems, which results in significant α variation; see Supplementary Section 2 for detailed discussion. A similar analysis is repeated for other regions, and μ values are shown in Figure 1f. For BL and TL, μ is also found to be nearly system-independent (Figure 1f). This suggests that sample-substrate intensity slopes can be used as a universal parameter to identify different layer-numbered regions accurately. Note that all the images are taken with objectives of similar numerical aperture (NA). Using objectives with different NA results in different intensity slopes[20,21] (shown in Supplementary Section 3, Table S1).

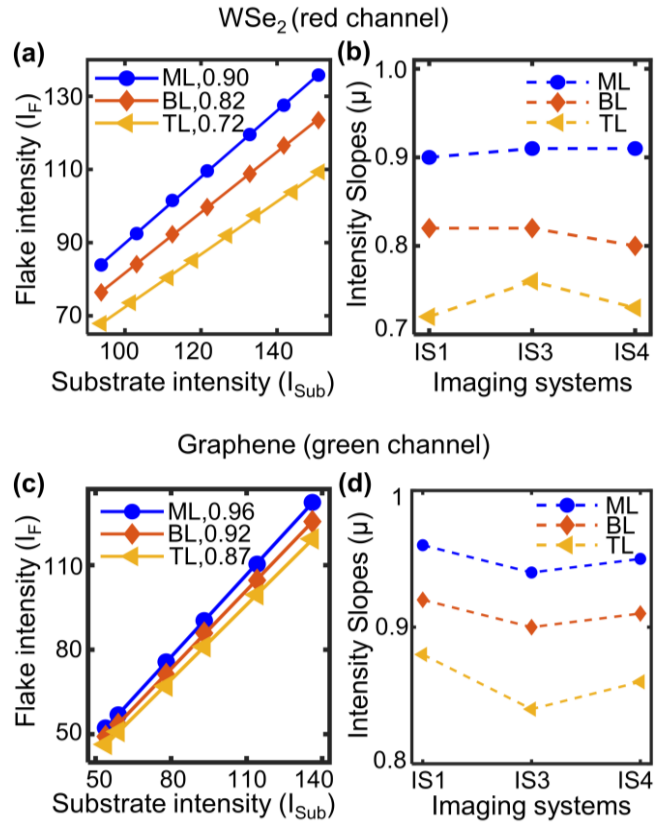

**Figure 2 (Intensity slope analysis for WSe$_2$ and graphene RGB images taken using different imaging systems): (a) & (c)** Flake intensity variation with substrate intensity, taken in RGB format, for WSe$_2$ and graphene flakes respectively. The variation is shown with linear fitting for each region. WSe$_2$ data uses red channel, and graphene uses green channel. **(b) & (d)** $\mu$ for red and green channels, for WSe$_2$ and graphene flakes respectively, measured using different imaging systems.

To demonstrate the robustness of our technique, we have performed slope determination method on mechanically exfoliated WSe$_2$ and graphene flakes. The optical images of the flakes and corresponding PL[22] and Raman spectrum[23] for layer identification are given in Supplementary Section 4. Similar to MoS$_2$, WSe$_2$ has the highest contrast for red channel images, whereas green channel is used for graphene[24–26]. Figure 2a shows the red channel $\mu$ for ML, BL and TL regions of WSe$_2$ flakes on 285nm SiO$_2$/Si substrate. $\mu$ measured using different imaging systems for WSe$_2$ is shown in figure 2b, which again demonstrates the nearly system-independent nature of intensity slopes. For graphene, we follow the same procedure using green channel images, and find different $\mu$ for regions with different layer number (ML, BL and TL, shown in Figure 2c). The difference between intensity slopes for different layer-numbered regions in graphene is not as high as for MoS$_2$ and WSe$_2$, which is possibly related to the lower thickness of graphene layers. Figure 2d again demonstrates the nearly system-independent measurement of intensity slopes, now in the case of graphene.

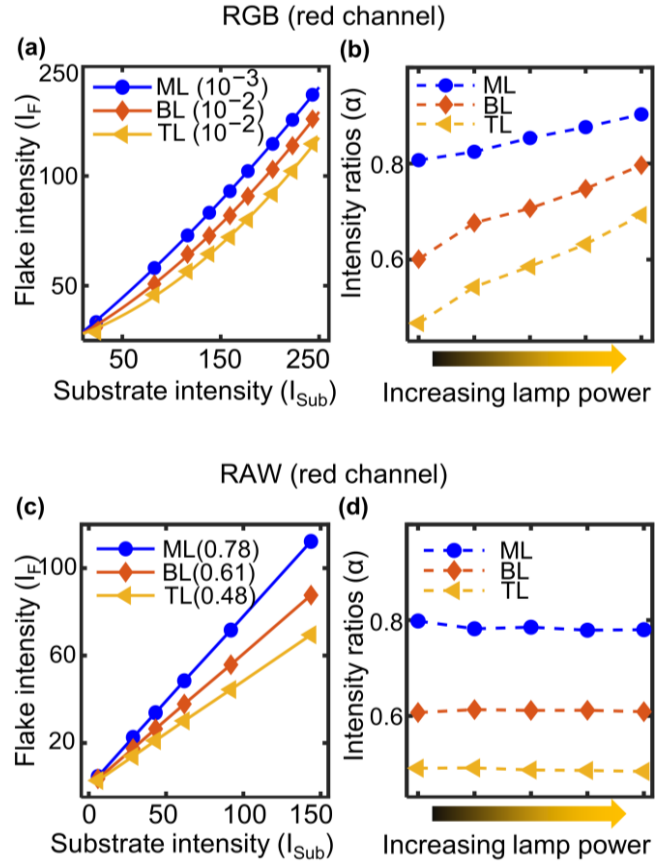

**Figure 3 (Intensity and slope comparison of RGB and RAW format imaging):** Variation of MoS$_2$ flake (same as in Figure 1) intensity with substrate intensity, for red channel images, taken in RGB & RAW format, respectively. **(a)** Variation for RGB format is shown with second order polynomial fitting ($P_1x^2 + P_2x + P_3$). $P_1/P_2$ values shown in the labels represent the magnitude of non-linearity in the variation. **(c)** Variation for RAW format is shown with linear fitting and slopes are mentioned in the labels. **(b) & (d)** The α for RGB and RAW format, respectively, with increasing lamp power. All images are taken using IS3.

Though $I_F$ vs. $I_{Sub}$ is linear in the thus far discussed intensity range (Figure 1e) for IS3, nonlinearities are observed when measuring a larger intensity range. Total intensity range variation is plotted in Figure 3a for ML, BL, and TL region of MoS$_2$ flake (same as Figure 1) for RGB. To illustrate the nonlinearity with lamp power variation, we have fitted the intensity values with a second order polynomial (solid curves in Figure 3a), ($P_1x^2 + P_2x + P_3$). The ratio of the second and first order coefficient ($P_1/P_2$) represents the magnitude of non-linearity of the intensity variations. The corresponding α is shown in Figure 3b with increasing lamp power (up to saturation of RGB pixel value, 255). This data is extracted from the images taken using IS3; similar dependence is found for IS4 (detailed comparison for all systems is given in Supplementary Section 2.). Both systems have the same camera and software but with different light sources (Table 1). The change in α is possibly related to the analog-to-digital conversion process of camera sensor data performed by the camera software, which includes compression, white balance and gamma correction. This processing enables the mapping to colour and intensity range of human vision, although this conversion is nonlinear[27,28].

To avoid these image processing-related effects, we took RAW format images that only record the intensity data collected by camera sensors[27,28]. Intensity plots extracted from RAW image format are shown in Figure 3c with linear best-fits. Remarkably, the intensity dependence is completely linear for the whole range of lamp power with nearly zero intercepts. The α values extracted from RAW images are plotted in Figure 3d, showing almost no change with increasing lamp power (data for IS4 is shown in Supplementary Figure S4). Thus, we reason that nonlinearity inherent to image processing, and non-zero intercept of $I_F$ vs. $I_{Sub}$ plots (see Supplementary Section 2), are the causes for the significant change of α with lamp power (in case of RGB images). The intensity slope is intensity and system independent and is truly a universal parameter to characterize layer number when using the RAW format. More importantly, intensity slope and ratio are same for RAW images, and hence α can directly be used to identify layer number. We thus propose that RAW format images are more reliable for measuring intensity slopes than the RGB format. However, we have discussed the RGB format due to the unavailability of RAW image capturing option in some imaging systems.

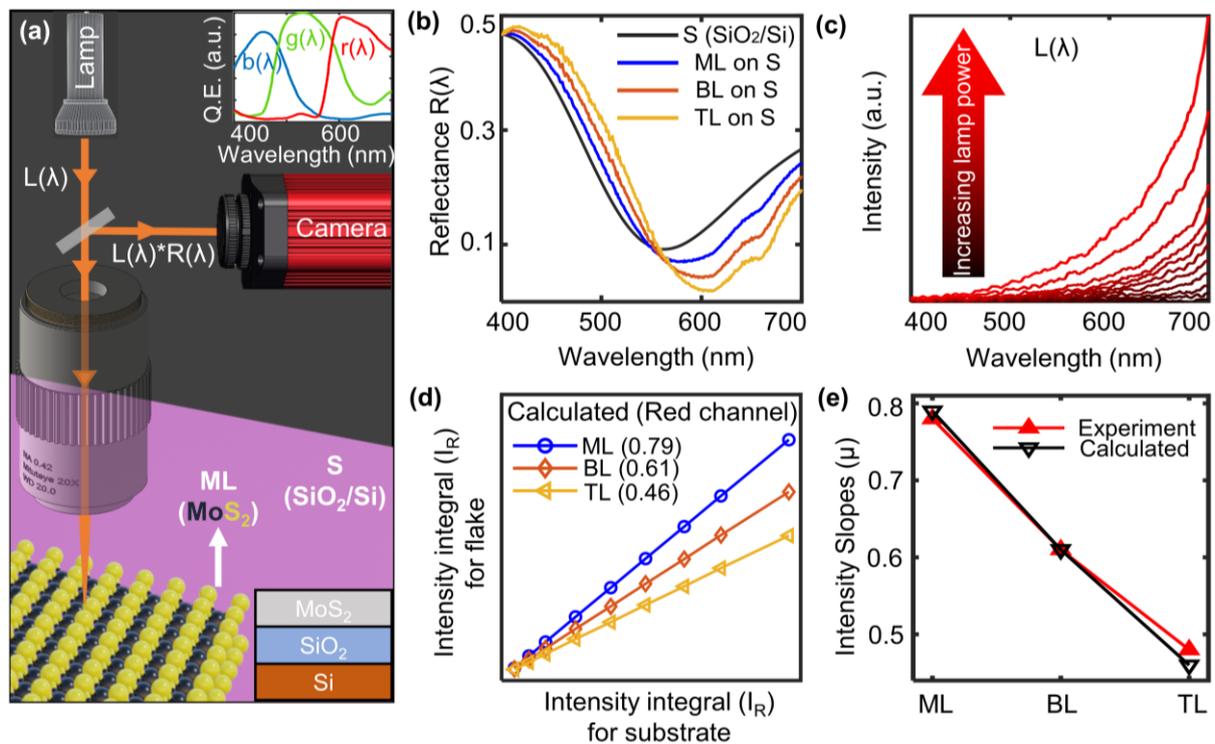

**Figure 4 (A Fresnel-reflectance-based optical model for calculation of the intensity slopes and comparison with experimentally measured values):** (a) Schematic of optical model used for calculation of intensity slopes. The top inset shows quantum efficiency (Q.E) curves for different channels of a typical CMOS camera sensor (shown data is for CS126CU colour CMOS camera sensor). The bottom inset is structural schematic of our system. (b) Calculated reflectance of 285nm $SiO_2$/Si substrate (black line), and ML, BL, and TL $MoS_2$ on substrate. (c) Halogen lamp spectrum measured by spectrometer (see methods) at different lamp powers (normalized with the highest intensity). (d) Calculated intensity integral ($I_R$) of reflected light for substrate, and ML, BL, and TL (on substrate), with increasing light intensity. Here, only red channel is shown at different lamp intensities (slopes are mentioned for each region in the legend, same as the ratio). (f) Comparison of calculated (black line) and experimentally measured (red line) µ values for RAW images.

We now propose an optical model to describe the imaging system. Figure 4a is the schematic of the model, indicating the major components of a microscope imaging setup, along with sample and substrate. Reflectance of a sample region depends on the thickness of the 2D material (number of layers), thickness of underlying $SiO_2$ layer (or other substrates) and corresponding materials' refractive indices. The Fresnel reflectance, calculated by incorporating the above parameters is shown in Figure 4b[29]. $MoS_2$, $SiO_2$ and Si refractive indices have been reliably measured in literature, and we directly use these values[16,30,31]. Spectrum of lamp ($L(\lambda)$) depends on the type of lamp (halogen, LED). Spectrum of halogen lamp (used in IS3) at different lamp powers is shown in Figure 4c. The reflected light spectrum ($R(\lambda)*L(\lambda)$) results in different colour for different regions. This reflected light gets detected by the camera sensor, with different sensitivities for red, green and blue channels (sensitivity curves of a typical CMOS camera are shown in the inset of Figure 4a). Thus, the integrated output of each channel can be expressed as an intensity integral, which represent the pixel values for different channels

$$I_R = \int R(\lambda) * L(\lambda) * r(\lambda) * d\lambda + c_r$$

$$I_G = \int R(\lambda) * L(\lambda) * g(\lambda) * d\lambda + c_g$$

$$I_B = \int R(\lambda) * L(\lambda) * b(\lambda) * d\lambda + c_b$$

where, c is dark counts of the system[32]. Here, c is taken to be zero for RAW images (non-zero for RGB). The integrated $I_R$ intensity values at different lamp powers are normalised with the maximum intensity value (substrate region at highest lamp power) and plotted in Figure 4d, for the different layer numbered regions. To compare the optical model with experiments, we used RAW image data (shown in Figure 3c) to avoid image-processing effects associated with RGB format[27,28]. The experimentally found intensity slopes are nearly identical to the calculated values, Figure 4e.

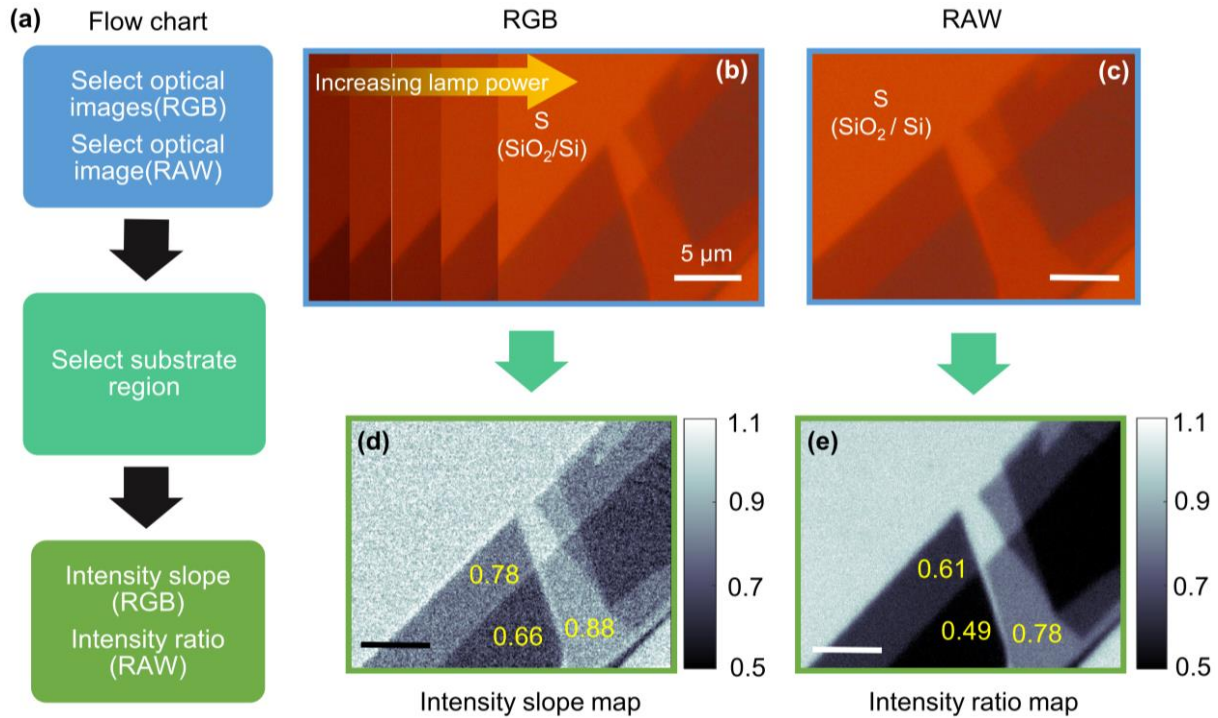

**Figure 5 (Graphical user interface (GUI) flowchart and examples for RGB and RAW images):** **(a)** Flowchart of the GUI. **(b)** and **(d)** RGB format: Multiple images (at different lamp powers) of $MoS_2$ flake on $SiO_2/Si$ substrate given as input in the GUI, output is slope map of the image. $S(SiO_2/Si)$ is the substrate region. Slopes of different regions are mentioned in the figure. **(c)** and **(e)** RAW format: One image is given as input, and output is a ratio map. The ratio of different regions is mentioned on the map. Both examples are for the red channel. All the scale bars are 5 μm.

To simplify the adoption of this technique for researchers, we have developed a graphical user interface (GUI) by using MATLAB (flowchart of the GUI is shown in Figure 5a), named SLOPEY. SLOPEY can be used to quickly analyse RGB (RAW) format images to check local slope (ratio) values, and to get a slope (ratio) map. For RGB format, images at multiple lamp powers are needed as input (figure 5b), whereas a single image is sufficient for the RAW format (figure 5c). Next, the substrate region must be selected for both cases. From the input, SLOPEY can produce a slope map for RGB format (figure 5d) and a ratio map for RAW format images (figure 5e). These slope and ratio values give direct confirmation of the layer number with high confidence.

**Conclusion section:**

In this study, we have developed a universal parameter which can be used for determination of layer number, irrespective of the microscope system, for conventionally used RGB and RAW formats. Also, we discuss why the intensity ratio (and $C_D$) is not a reliable parameter for RGB format, whereas it suffices for RAW format. For RGB, intensity slope is found to be linear for the lamp power range usually used to image a sample. For some systems where the intensity variations are not linear throughout the intensity range (like S3 and S4), the RAW format can be an excellent solution that gives completely system and intensity-independent slope values. We also calculate slopes and ratios using a Fresnel-reflectance model and verify our measurements. We have developed a GUI, where we create a slope (or ratio) map for a specified region of interest. Through this study we have satisfied the longstanding need for a confirmatory universal optical contrast method.

This slope method and the provided GUI do not need any special system equipment, making it ready to be adopted in any lab with an optical microscope. We believe this process can be extended to any 2D materials. Database of slope and ratio values can be created for different 2D materials on different substrates. This method is expected to accelerate the identification of layer number and reduce the fabrication time of heterostructures.

**Methods:**

Sample preparation: $MoS_2$ and $WSe_2$ bulk crystals are purchased from 2D Semiconductors, graphene from NGS Naturgraphit GmbH. $MoS_2$, $WSe_2$ and graphene bulk crystals are exfoliated using conventional scotch tape method. We transferred graphene flakes directly to the substrate from the scotch tape. For the TMDs, we have used cell phone PET film screen protectors to transfer the flakes.

Optical measurements: Flakes on $SiO_2$/Si substrate are imaged using different optical imaging systems, list is given in Table 1. The spectrum of halogen lamp at different lamp powers are taken using Kymera-328iB1 spectrometer equipped with Andor iDUS-416 CCD. All Raman and PL measurements are done using 532 nm laser in a Horiba LabRAM HR setup.

**Author contributions**

MM and AS developed the experimental and theoretical framework. MM performed the optical experiments and mechanical exfoliation, with assistance from AKD. MM performed the data analysis. MM and AS discussed and prepared the manuscript, with contributions from AKD.

**Acknowledgments**

AS would like to acknowledge funding from Indian Institute of Science start-up grant. AKD would like to acknowledge Prime Minister's Research Fellowship (PMRF).

**Data availability**

All data is available upon reasonable request. The computer code concerning the MATLAB GUI has been uploaded at https://github.com/ousumsPhysics/SLOPEY .

**References:**

1. Wang, G. *et al.* Colloquium: Excitons in atomically thin transition metal dichalcogenides. *Rev. Mod. Phys.* **90**, 021001 (2018).

2. Tran, K., Choi, J. & Singh, A. Moiré and beyond in transition metal dichalcogenide twisted bilayers. *2D Mater.* **8**, 022002 (2020).

3. Lopez-Sanchez, O., Lembke, D., Kayci, M., Radenovic, A. & Kis, A. Ultrasensitive photodetectors based on monolayer MoS2. *Nature Nanotech* **8**, 497–501 (2013).

4. Yan, J. *et al.* Dual-gated bilayer graphene hot-electron bolometer. *Nature Nanotechnology* **7**, 472–478 (2012).

5. Gottscholl, A. *et al.* Spin defects in hBN as promising temperature, pressure and magnetic field quantum sensors. *Nat Commun* **12**, 4480 (2021).

6. He, Y.-M. *et al.* Single quantum emitters in monolayer semiconductors. *Nature Nanotech* **10**, 497–502 (2015).

7. Parto, K., Azzam, S. I., Banerjee, K. & Moody, G. Defect and strain engineering of monolayer WSe2 enables site-controlled single-photon emission up to 150 K. *Nat Commun* **12**, 3585 (2021).

8. Regan, E. C. *et al.* Mott and generalized Wigner crystal states in WSe2/WS2 moiré superlattices. *Nature* **579**, 359–363 (2020).

9. Baek, H. *et al.* Optical read-out of Coulomb staircases in a moiré superlattice via trapped interlayer trions. *Nat. Nanotechnol.* **16**, 1237–1243 (2021).

10. Ottaviano, L. *et al.* Mechanical exfoliation and layer number identification of MoS 2 revisited. *2D Mater.* **4**, 045013 (2017).

11. Huang, J.-K. *et al.* Large-Area Synthesis of Highly Crystalline WSe2 Monolayers and Device Applications. *ACS Nano* **8**, 923–930 (2014).

12. Bing, D. *et al.* Optical contrast for identifying the thickness of two-dimensional materials. *Optics Communications* **406**, 128–138 (2018).

13. Li, H. *et al.* Rapid and Reliable Thickness Identification of Two-Dimensional Nanosheets Using Optical Microscopy. *ACS Nano* **7**, 10344–10353 (2013).

14. Li, H. *et al.* Optical Identification of Single- and Few-Layer MoS2 Sheets. *Small* **8**, 682–686 (2012).

15. Li, Y. *et al.* Rapid identification of two-dimensional materials via machine learning assisted optic microscopy. *Journal of Materiomics* **5**, 413–421 (2019).

16. Zhang, H. *et al.* Measuring the Refractive Index of Highly Crystalline Monolayer MoS 2 with High Confidence. *Sci Rep* **5**, 8440 (2015).

17. Li, Y. *et al.* Optical identification of layered MoS2via the characteristic matrix method. *Nanoscale* **8**, 1210–1215 (2015).

18. Lee, C. *et al.* Anomalous Lattice Vibrations of Single- and Few-Layer MoS2. *ACS Nano* **4**, 2695–2700 (2010).

19. Li, H. *et al.* From Bulk to Monolayer MoS2: Evolution of Raman Scattering. *Advanced Functional Materials* **22**, 1385–1390 (2012).

20. Saigal, N., Mukherjee, A., Sugunakar, V. & Ghosh, S. Angle of incidence averaging in reflectance measurements with optical microscopes for studying layered two-dimensional materials. *Review of Scientific Instruments* **85**, 073105 (2014).

21. Gao, L., Ren, W., Li, F. & Cheng, H.-M. Total Color Difference for Rapid and Accurate Identification of Graphene. *ACS Nano* **2**, 1625–1633 (2008).

22. Li, Y. *et al.* Accurate identification of layer number for few-layer WS2 and WSe2 via spectroscopic study. *Nanotechnology* **29**, 124001 (2018).

23. Graf, D. *et al.* Spatially Resolved Raman Spectroscopy of Single- and Few-Layer Graphene. *Nano Lett.* **7**, 238–242 (2007).

24. Wang, Y. Y. *et al.* Thickness identification of two-dimensional materials by optical imaging. *Nanotechnology* **23**, 495713 (2012).

25. Ermolaev, G. A. *et al.* Spectroscopic ellipsometry of large area monolayer WS2 and WSe2 films. *AIP Conference Proceedings* **2359**, 020005 (2021).

26. Weber, J. W., Calado, V. E. & van de Sanden, M. C. M. Optical constants of graphene measured by spectroscopic ellipsometry. *Appl. Phys. Lett.* **97**, 091904 (2010).

27. Schewe, J. *The Digital Negative: Raw Image Processing in Lightroom, Camera Raw, and Photoshop*. (Pearson Education, 2015).

28. Andrews, P., Butler, Y., Butler, Y. J. & Farace, J. *Raw Workflow from Capture to Archives: A Complete Digital Photographer's Guide to Raw Imaging*. (Focal Press, 2006).

29. Blake, P. *et al.* Making graphene visible. *Appl. Phys. Lett.* **91**, 063124 (2007).

30. Aspnes, D. E. & Studna, A. A. Dielectric functions and optical parameters of Si, Ge, GaP, GaAs, GaSb, InP, InAs, and InSb from 1.5 to 6.0 eV. *Phys. Rev. B* **27**, 985–1009 (1983).

31. Gao, L., Lemarchand, F. & Lequime, M. Exploitation of multiple incidences spectrometric measurements for thin film reverse engineering. *Opt. Express, OE* **20**, 15734–15751 (2012).

32. Jessen, B. S. *et al.* Quantitative optical mapping of two-dimensional materials. *Sci Rep* **8**, 6381 (2018).

# Supplementary information for

## Optical microscope based universal parameter for identifying layer number in two-dimensional materials

Mainak Mondal, Ajit Kumar Dash, Akshay Singh*

Department of Physics, Indian Institute of Science, Bengaluru, India- 560012

Corresponding author- *aksy@iisc.ac.in*

**Section 1:**

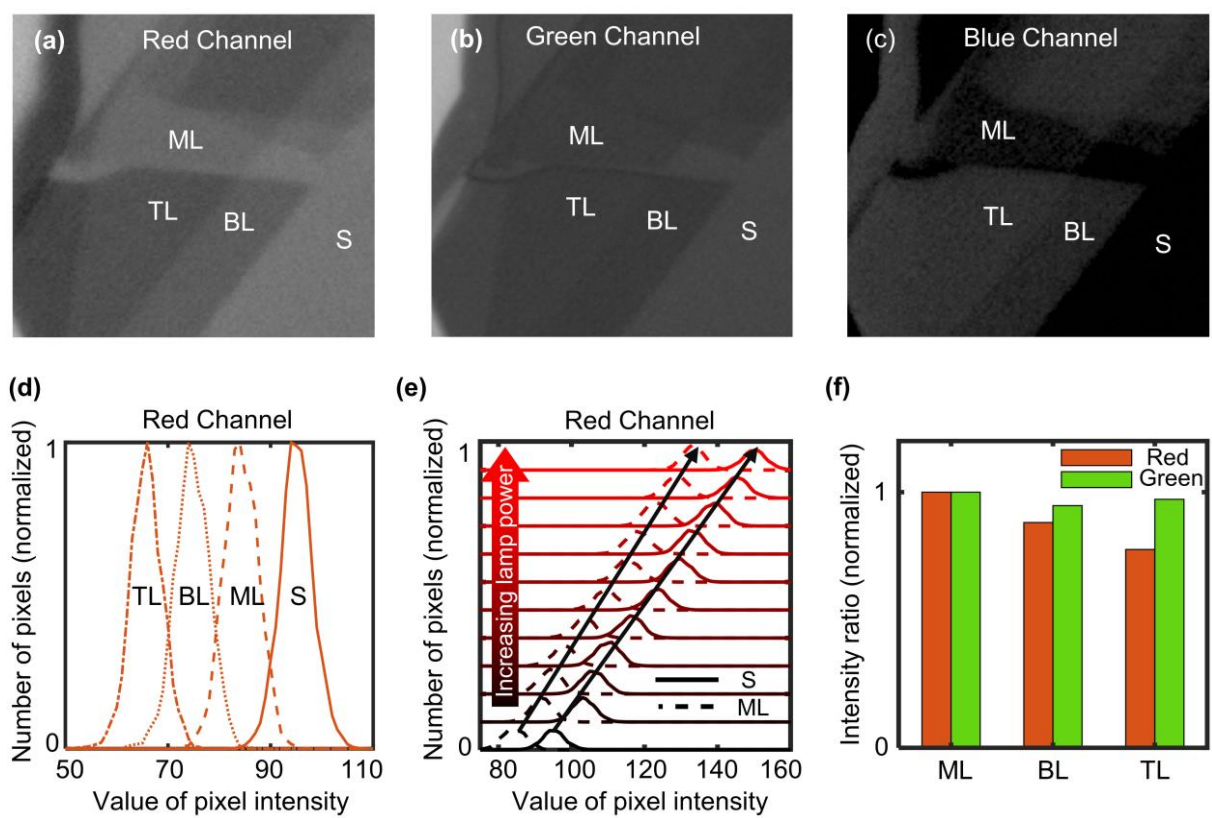

**Figure S1 (Extraction of mean intensity values of different regions from the images): (a), (b) & (c)** Gray scale images of MoS$_2$ flake on 285nm SiO2/Si substrate extracted from red, green, and blue channel respectively. Different regions ML, BL, TL of the flake, and substrate(S) are labelled on the images. **(d)** Normalized histograms of pixel values distribution for different regions are shown, corresponding to the red channel image. **(e)** Comparison of distribution histograms with increasing lamp power for the substrate and ML region are shown. Mean value of the distribution is seen to change at different rates, indicated by two arrows. **(f)** The intensity ratio of ML, BL and TL regions for red, green channel images. For blue channel, the substrate intensity is close to zero and hence the ratio is not shown.

In Figure 1a in the main text, a RGB format image of the MoS$_2$ flake is shown. In this format each pixel is given a red, green, and blue colour shade depending on the object colour which is being imaged. There are 256 shades in a standard RGB image for each colour. Next, red channel gray scale image is extracted using MATLAB (shown in Figure S1a), where each pixel only consists of red colour shade. Similarly gray scale images for green and blue channels are shown in Figure S1b and S1c. We focused on red channel because this channel provides consistent intensity variation for different regions (Figure S1f). The mean shade value for a particular region is the reflected intensity for that region. To calculate the mean shade value, first each region is cropped out from the total image. The histogram of pixel intensity values and corresponding number of pixels with that value are shown in Figure S1d for substrate, ML, BL, and TL regions. The mean of pixel intensity value of each distribution gives the corresponding substrate intensity ($I_{Sub}$) or flake intensity ($I_F$). Comparison of the same histogram for the images taken at different lamp intensities are shown in Figure S1e (ML and Substrate). With increasing lamp power, shifting of distribution can be noticed for both region's distribution, but at different rates. The mean value from each intensity is then plotted for calculating the slopes as shown, in Figure 1d of the main text.

**Section 2:**

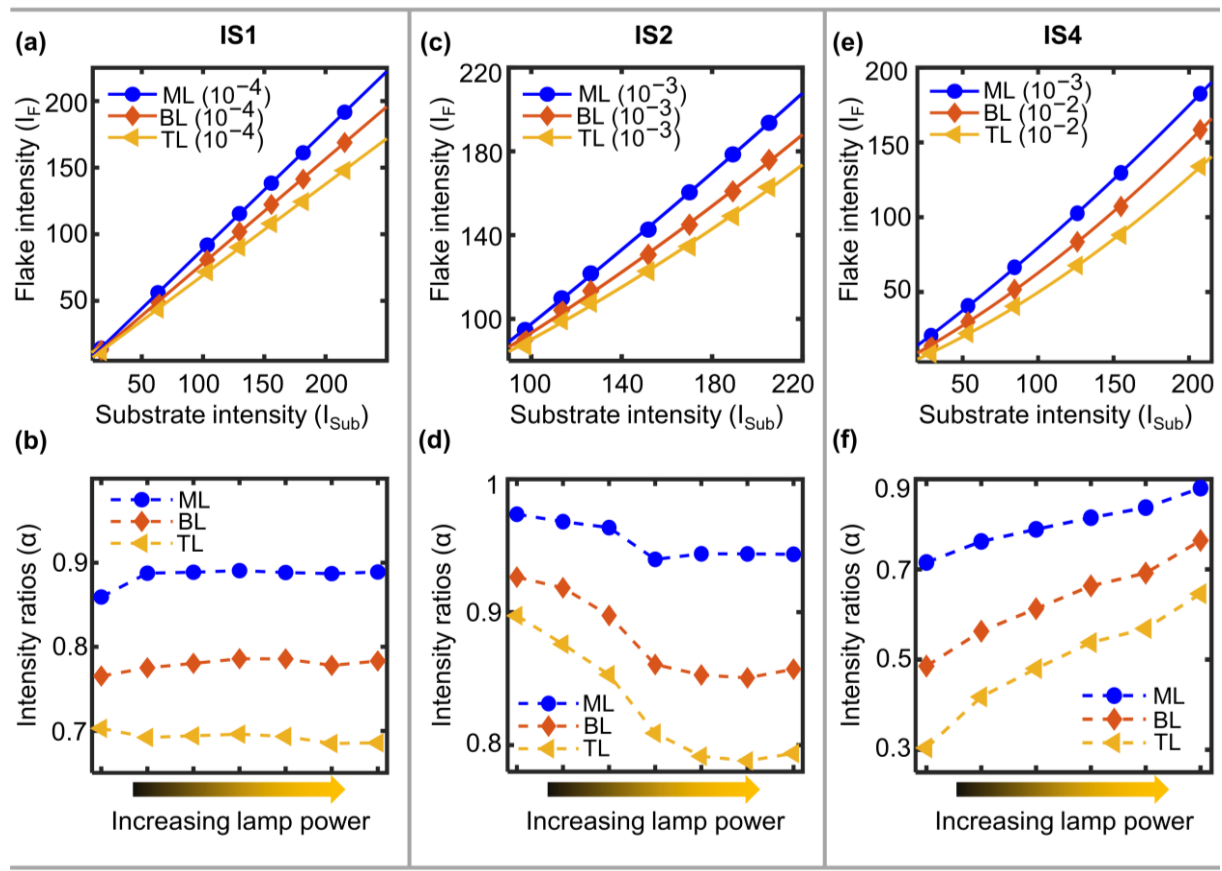

**Figure S2 (Large range intensity and ratio variation for IS1, IS2 and IS4):** For IS1, **(a)** full range variation of the reflected flake (ML, BL, TL) intensity ($I_F$) with substrate intensity ($I_{Sub}$) for increasing lamp light power. Solid lines are second order polynomial fits ($P_1x^2 + P_2x + P_3$), $P_1/P_2$ values shown in the labels represent the non-linearity in the variation. **(b)** Intensity ratio (α) variation with increasing lamp power for different regions. For IS2, **(c)** large range $I_F$ vs. $I_{Sub}$ variation with increasing lamp power. Solid lines are second order polynomial fit. **(d)** Intensity ratio (α) variation with increasing lamp power for different regions. For IS4, **(e)** large range $I_F$ vs. $I_{Sub}$ variation with increasing lamp power. Solid lines are second order polynomial fit. **(f)** Intensity ratio (α) variation with increasing lamp power for different regions.

Figure S2 shows large range intensity variation (S2 a,c,e) and corresponding ratios (S2 b,d,f) for RGB format images of different regions taken using IS1, IS2 and IS4. Variation for IS3 is shown in Figure 3a, main text. Solid lines represent the second order polynomial fitting ($P_1x^2 + P_2x + P_3$), and $P_1/P_2$ ratio corresponds to the non-linearity of each variation. IS1 shows mostly linear behaviour, whereas for IS3 and IS4 significant non-linearity is present. Non-linearity also increases with increasing layer number.

For linear dependence, $I_F = a*I_{Sub} + b$, where a is the slope and b is the intercept. Following this, $\alpha = a + \frac{b}{I_{Sub}}$. This suggests for b ~ 0, α will be independent of $I_{Sub}$ and will be constant. However, for b ≠ 0, α will have a dependence on $I_{Sub}$. For RGB format, the intercept values for the linear fitting (for ML) shown in Figure 1e are 0.02, 9.51, -8.61 and -9.44 for IS1, IS2, IS3 and IS4 respectively. Variation of α (for ML) for different systems are shown in Figure S2b, d, f and 3b with changing lamp power. Lower intercepts result in almost no changes throughout large intensity variation, whereas higher intercepts show large changes in α. In case of RAW format, the intercepts are zero and as a result there is no variation in intensity ratio parameter with changing light intensity.

**Section 3:**

| Numerical aperture | ML | BL | TL |
|---|---|---|---|
| 0.5 (20X objective) | 0.83 | 0.68 | 0.55 |
| 0.8 (50X objective) | 0.89 | 0.77 | 0.67 |
| 0.9 (100X objective) | 0.91 | 0.81 | 0.73 |

**Table S1:** Regional Slope variation with different objectives for different layer numbered regions.

Images of the MoS$_2$ flake (shown in Figure 1a) taken using IS4 using three different objectives with numerical aperture (NA) 0.5, 0.8, and 0.9. All other imaging conditions are kept same.

Calculated slopes for ML, BL and TL regions are shown in the Table 1. Significant variation is observed for imaging with different objectives. The reason for such variation is the change of light collection angle as we change the NA. For lower NA objectives mostly perpendicular reflected light is collected, whereas for higher NA off-axis light also reaches the sensor. As the reflection coefficient of the sample varies with incident angle, different NA objectives show different colour variation for the same region, and results in variation of intensity slopes with NA[1,2].

**Section 4:**

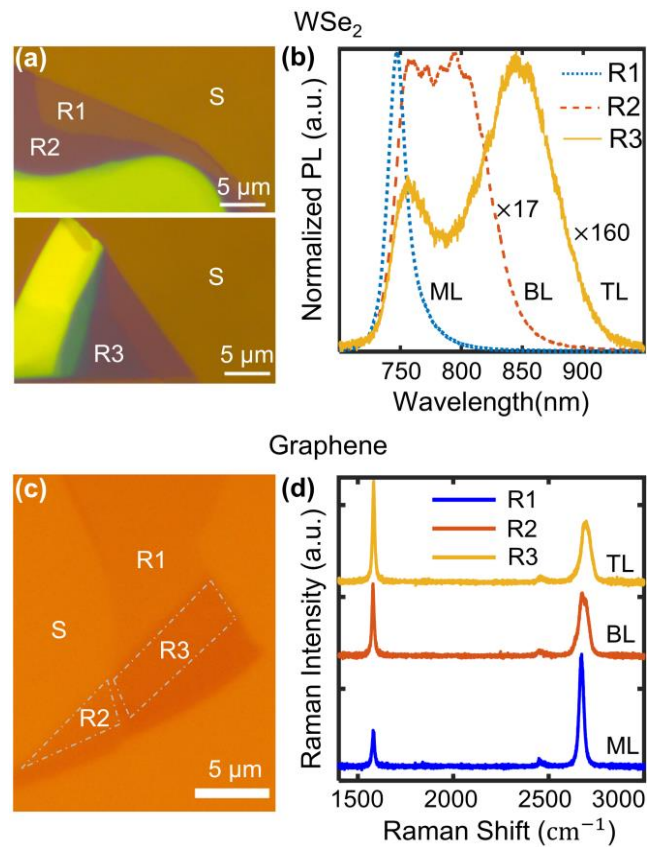

**Figure S3 (PL and Raman of WSe$_2$ and graphene): (a)** Flakes of WSe$_2$, different regions are named as R1, R2, R3 and S, images are taken using IS1. S is the substrate. Scale bars indicate 5μm. **(b)** The PL of corresponding flake regions. PL confirms the three regions are ML, BL and TL respectively[3]. **(c)** The graphene flake regions are named similarly (as WSe$_2$ regions), image is taken using IS3. Scale bar indicates 5μm. **(d)** Raman of each region is shown in S3d, confirming the layer number[4].

**Section 5:**

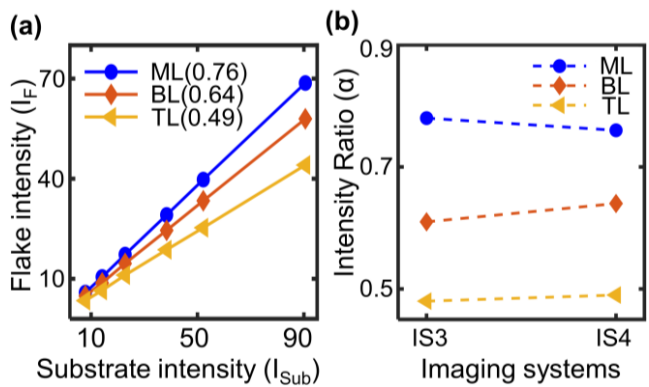

**Figure S4 (RAW channel data for IS4 and intensity ratio comparison for different imaging systems):** **(a)** Variation of $MoS_2$ flake (same as in Figure 1) intensity with substrate intensity for red channel images taken in RAW format using IS4. The variation is shown with first order polynomial fitting for each layered region. **(b)** Intensity ratio comparison for IS3 and IS4 for different layered regions of $MoS_2$.

**Section 6:**

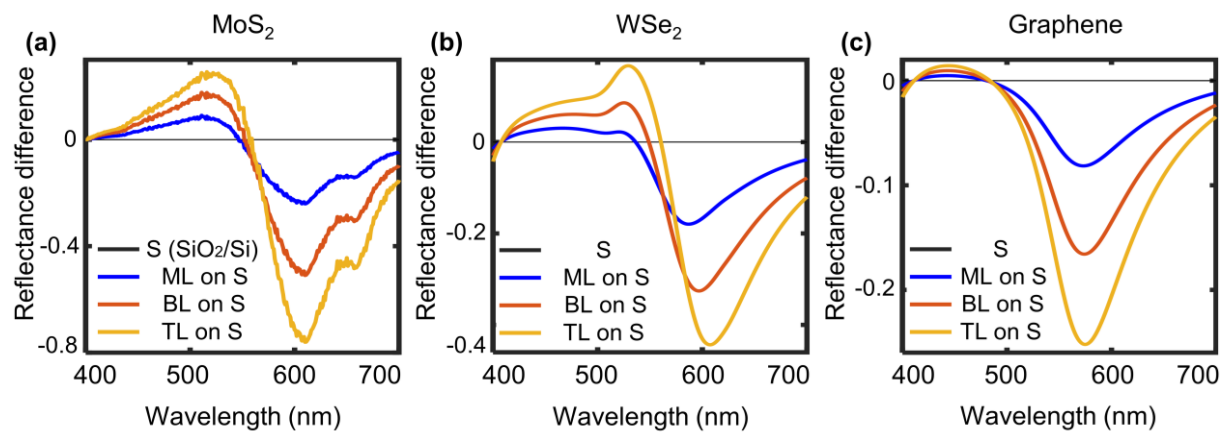

**Figure S5 (Calculated reflectance difference between the flake and substrate for ML, BL and TL regions):** **(a), (b)** and **(c)** Calculated reflectance difference (Reflectance difference $= \frac{\text{(Flake reflectance - Substrate reflectance)}}{\text{(Flake reflectance + Substrate reflectance)}}$) for $MoS_2$, $WSe_2$ and graphene, respectively. Calculation done for 285nm $SiO_2$/Si substrate[5–10].

**References:**

1. Saigal, N., Mukherjee, A., Sugunakar, V. & Ghosh, S. Angle of incidence averaging in reflectance measurements with optical microscopes for studying layered two-dimensional materials. *Review of Scientific Instruments* **85**, 073105 (2014).

2. Gao, L., Ren, W., Li, F. & Cheng, H.-M. Total Color Difference for Rapid and Accurate Identification of Graphene. *ACS Nano* **2**, 1625–1633 (2008).

3. Li, Y. *et al.* Accurate identification of layer number for few-layer WS2 and WSe2 via spectroscopic study. *Nanotechnology* **29**, 124001 (2018).

4. Graf, D. *et al.* Spatially Resolved Raman Spectroscopy of Single- and Few-Layer Graphene. *Nano Lett.* **7**, 238–242 (2007).

5. Li, Y. *et al.* Rapid identification of two-dimensional materials via machine learning assisted optic microscopy. *Journal of Materiomics* **5**, 413–421 (2019).

6. Zhang, H. *et al.* Measuring the Refractive Index of Highly Crystalline Monolayer MoS 2 with High Confidence. *Sci Rep* **5**, 8440 (2015).

7. Ermolaev, G. A. *et al.* Spectroscopic ellipsometry of large area monolayer WS2 and WSe2 films. *AIP Conference Proceedings* **2359**, 020005 (2021).

8. Aspnes, D. E. & Studna, A. A. Dielectric functions and optical parameters of Si, Ge, GaP, GaAs, GaSb, InP, InAs, and InSb from 1.5 to 6.0 eV. *Phys. Rev. B* **27**, 985–1009 (1983).

9. Gao, L., Lemarchand, F. & Lequime, M. Exploitation of multiple incidences spectrometric measurements for thin film reverse engineering. *Opt. Express, OE* **20**, 15734–15751 (2012).

10. Weber, J. W., Calado, V. E. & van de Sanden, M. C. M. Optical constants of graphene measured by spectroscopic ellipsometry. *Appl. Phys. Lett.* **97**, 091904 (2010).